\documentclass[aps,prl,twocolumn,floatfix,longbibliography]{revtex4-1}
\usepackage{enumitem}
\usepackage{graphicx}
\usepackage{bm}
\usepackage{epsf}
\usepackage{ulem}
\usepackage[colorlinks = true]{hyperref}
\usepackage{amssymb}
\usepackage{amsmath}
\usepackage{graphicx}
\usepackage{rotating}
\usepackage{epsfig}
\usepackage{psfrag}
\usepackage{amsmath}
\usepackage{subfigure}

\newcommand{\gtappr}{{{\lower4pt\hbox{$>$} } \atop \widetilde{ \ \ \ }}}

\DeclareGraphicsExtensions{.pdf}
\newlength{\figwidth}
\newcommand{\fg}[3]
{\begin{figure}[htb]\vspace*{-0cm}\centerline{\includegraphics[width=\figwidth]{#1}}\vskip
-0.2cm \caption{\label{#2}#3}\end{figure}}
\newcommand{\urs}{URu$_{2}$Si$_{2}$\ }
\newcommand{\urss}{URu$_{2}$Si$_{2}$}

\begin{document}
\title{Implications of the Measured Angular Anisotropy at the Hidden Order Transition of \urs}
%
\author{P. Chandra,$^{1}$ P. Coleman,$^{1,2}$ R. Flint,$^{3}$ J. Trinh$^{4}$
and A. P. Ramirez$^{4}$}
\affiliation{$^1$Center for Materials Theory, Rutgers University, Piscataway, New Jersey, 08854, USA \\ 
$^2$Department of Physics, Royal Holloway, University of London, Egham, Surrey TW20 0EX, UK}
\affiliation{$^3$Department of Physics and Astronomy,
	Iowa State University, Ames, Iowa, 50011}
\affiliation{$^4$Physics Department,  UC Santa Cruz, 
Santa Cruz, California, 95064, USA}

\date{\today}

\begin{abstract}
\centerline{Contributed paper to SCES 2017, Prague.} \vskip 0.1in

The heavy fermion compound \urs continues to attract great interest
due to the long-unidentified nature of the hidden order that develops
below 17.5K. Here we discuss the implications 
of an angular survey of the linear and nonlinear
susceptibility of \urs
in the vicinity of the hidden order transition \cite{Trinh16}.
While the anisotropic nature of spin
fluctuations and low-temperature quasiparticles was previously
established, our recent results 
suggest that the order parameter itself has intrinsic
Ising anisotropy, and that moreover this anisotropy extends far above the
hidden order transition. Consistency checks and subsequent questions for 
future experimental and theoretical studies of hidden order are discussed. 
\end{abstract}
\maketitle

\section{Motivation}


Consensus has not been reached on the nature of the ``hidden order'' (HO) in \urs 
despite several decades of active research. At $T_{HO}= 17.5 K$ 
sharp features in the thermodynamic quantities accompanied by significant 
entropy loss ($S > \frac{1}{3} R \ln 2$), but to  date
no associated charge or spin ordering has 
been directly detected at ambient pressure \cite{Mydosh:2011wm}.
The nature of the quasiparticle excitations and the broken symmetries 
associated with the HO phase are important questions for understanding not 
only HO but also the exotic superconductivity that develops at low temperatures.

%

Several measurements on \urs indicate the importance of Ising anisotropy
in the HO phase despite the absence of local moments at these temperatures
and pressures.  At $T_{HO}$,  both the linear ($\chi_1$) 
and the nonlinear
($\chi_3$) susceptibilities are anisotropic, with $\chi_3$ in the easy
axis direction displaying
a sharp anomaly $\Delta \chi_3 = \chi_{3} (T_{c}^{-})-\chi_{3}
(T_{c}^{+})$ that tracks closely with the
structure of the specific heat \cite{Miyako91,Ramirez92}.
At lower temperatures, non-spinflip ($\Delta J_z =0 $) magnetic excitations
detected by inelastic neutron scattering \cite{Broholm91}
have Ising character.  Quantum oscillations measured deep within 
the HO region indicate a strongly anisotropic quasiparticle g-factor 
$g(\theta) \propto \cos \theta$, where $\theta$ is the
angle away from the c-axis \cite{Ohkuni:1999ig,Altarawneh:2011hia}.
This $g(\theta)$ is confirmed by upper critical field
experiments \cite{Altarawneh:2012cy}, 
indicating that heavy Ising
quasiparticles pair to form a Pauli-limited superconductor at low temperatures.

It is thus natural to ask whether the Ising nature of the itinerant
quasiparticles has its origin at $T_{HO}$.
Support for this idea 
is suggested by the observation that the Ising
anisotropy obtained from dHvA and the
superconducting upper-critical field
measurements \cite{Ohkuni:1999ig,Altarawneh:2011hia} far exceeds the five-fold
anisotropy seen in the bulk magnetic susceptibility \cite{Miyako91,Ramirez92}. 
However, to confirm this idea, 
another measurement is needed to probe the quasiparticle g-factors in 
the vicinity of the hidden order transition.  

\section{Angular Survey of the Hidden Order Transition with a Bulk Measurement}
%

\figwidth=\columnwidth
\fg{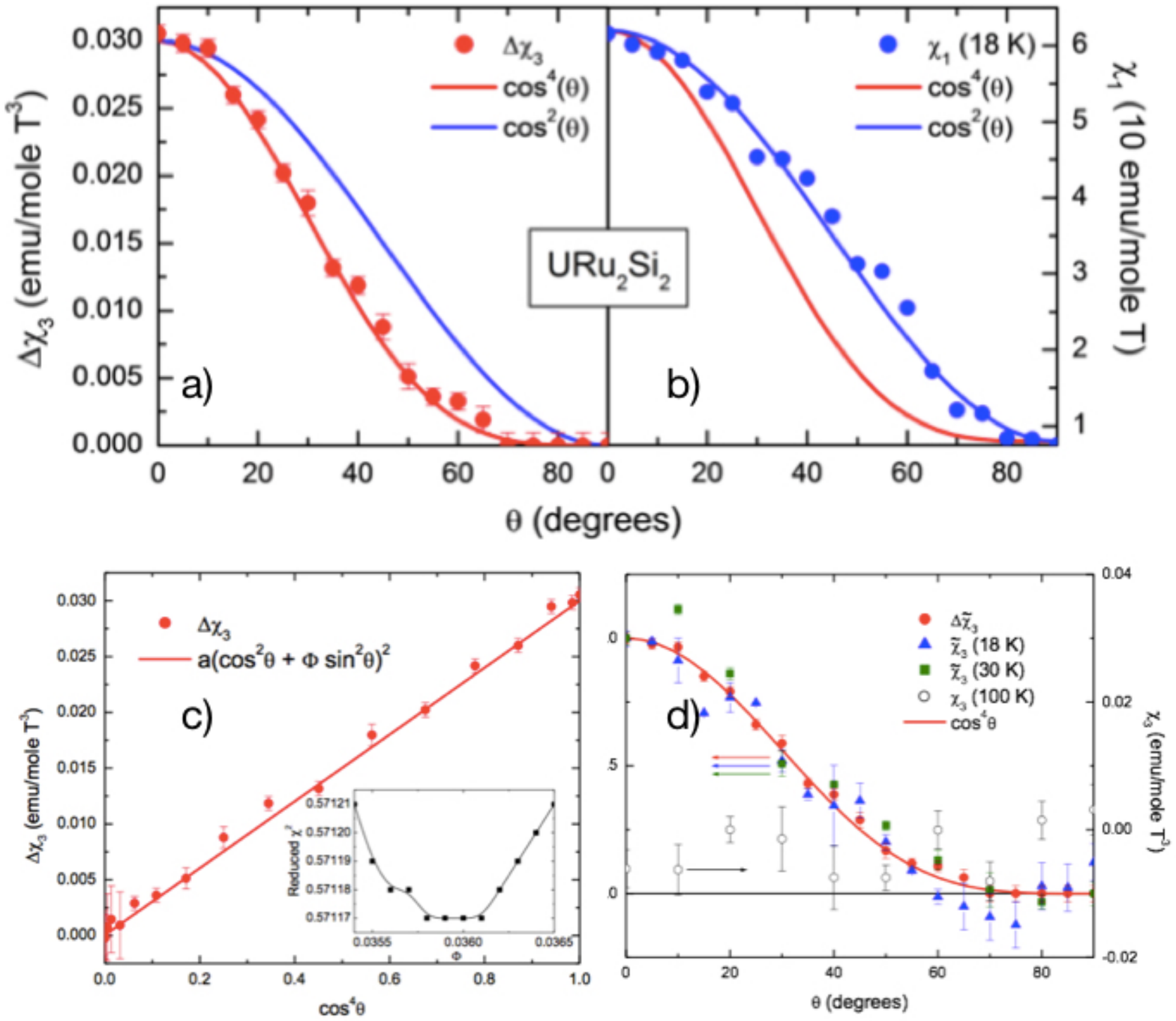}{SCES17_Figure1}{Angular dependence of a) $\Delta \chi_3$ and
b) $\chi$. c) Fidelity of Ising anisotropy fit and 
d) $\chi_3$ for temperatures above $T_c$. Figure adapted from \cite{Trinh16}.}

The general expression for the field-dependent part of the free
energy in a tetragonal crystal at fixed temperature is
\begin{equation}
F = - \chi_1 (\theta) \frac{H^2}{2} - \chi_3 (\theta,\phi) \frac{H^4}{4!} 
\end{equation}
where $\theta$ and $\phi$ refer to the 
angles away from the c-axis and in the basal plane respectively and 
details of this angular decomposition can be found elsewhere \cite{Trinh16};
for simplicity here we take $\mu_0 = 1$, so that  $\mu_{0}H\equiv H$
is the external field, measured in Tesla. 
Because $\Delta \chi_3$ \cite{Ramirez91} is determined by the
excitations near the Fermi surface, it is ideally suited as 
a direct thermodynamic probe of the electronic g-factors at 
the HO transition \cite{Trinh16}.   
Consistency with the low-temperature $g(\theta) \propto \cos \theta$
results \cite{Ohkuni:1999ig,Altarawneh:2011hia,Altarawneh:2012cy}, 
requires a $\Delta \chi_3 (\theta) \propto \cos^4\theta$ since 
$\Delta \chi_3 (\theta) \propto \{g(\theta)\}^4$ 
\cite{Chandra:2013gv,Chandra14,Chandra15}. 

Details of the linear and the nonlinear susceptibility measurements as 
a function of angle can be found elsewhere and here we simply
summarize the main results \cite{Trinh16}.   In Figure 1 the
angular dependence of $\Delta \chi_3$ and of $\chi$ at and just above the HO 
transition is presented. 
The linear susceptibility displayed in figure 1b is characterized
by the form
\begin{equation}
\chi_1(\theta,T) = \chi_1^{(0)} + \chi_1^{Ising} (T) \cos^2 \theta,
\end{equation}
where the isotropic (Van Vleck) component $\chi_1^{(0)}$ of the susceptibility 
displays no discernible temperature dependence. 
Whereas $\chi_1 (\theta)$ varies as $\cos^2 \theta$
at $T = 18 K$, in Fig. 2 a)  
$\Delta \chi_3$ has a distinctive $\cos^4\theta$ dependence 
\begin{equation}
\Delta\chi_3 (\theta,\phi) = \Delta \chi_3^{Ising} \cos^4 \theta
\end{equation}
without any Van Vleck (constant) terms, 
consistent with the low-temperature $g(\theta)$ measurements. In Figure 2c
the robustness of the Ising anisotropy is codified \cite{Trinh16} 
by considering an angle-dependent coupling between the hidden order 
parameter and the magnetic field that results in
\begin{equation}\label{fitting}
\Delta \chi_{3} (\theta ) \propto (\cos^2\theta+ \Phi \sin^2 \theta )^{2}.
\end{equation}  
where $\Phi$ quantifies the fidelity of Ising behavior.
Our measurements indicate a very small $\Phi = 0.036\pm0.021$, 
shown in figure 2c (inset), that could be due to an 
angular offset of only one degree; details of the fitting procedure can be 
found elsewhere \cite{Trinh16}.

These results, at the very least, indicate that the free energy 
of \urs only depends on the $z$ component of the
magnetic field, namely $F[\vec{H}] = F[H_{z}]$.  This in turn
implies that the Zeeman term in the microscopic Hamiltonian 
$H_{Zeeman} \propto -J_{z}B_{z}$ is coupled to the 
single-ion properties of the $U$ ions in \urs via 
hybridization with the conduction
electrons. The observed
Ising anisotropy also suggests an integer
spin $5f^2$ U ground-state. This point of view is further supported 
by both dynamical mean-field theory
\cite{Haule09} and high-resolution RIXs measurements
\cite{Wray:2015gga}.

However this picture is incomplete, for 
the sharpness of the specific-heat anomaly, the
sizable entropy and the gapping of two-thirds of the Fermi surface
associated with the hidden order transition\cite{Mydosh:2011wm}
indicate an underlying
{\sl itinerant} ordering process, as if the hybridization itself is
the order parameter\cite{Chandra:2013gv}.  An intriguing feature of
these results
is that the jump $\Delta \chi_{3}$ that reflects the itinerant
ordering process also exhibits a strong Ising anisotropy. 
Furthermore, as shown in Figure 3d, there is a positive anisotropic
$\chi_3$ to temperatures well above the hidden order transition and this
cannot be explained with single-ion physics. We therefore must consider
the strong Ising character of the hidden order parameter, now that 
we have established that we have heavy Ising quasiparticles at the hidden
order transition.  The reconciliation
of the single-ion and the itinerant perspectives, both supported by experiment, presents a fascinating challenge in \urs. 

\section{Consistency Checks}

It is important to cross-check these susceptibility results with other experimental measurements on \urss.
At the HO transition, our results can be analyzed within a minimal Landau free
energy density of the form 
\begin{equation}
f[T,\psi] = a [T-T_c ({\bf H})]\psi^2 + \frac{b}{2}\psi^4,
\label{fenergy}
\end{equation}
where $\psi $ is the hidden order parameter, we continue to take $\mu_0 = 1$ for simplicity, and 
\begin{equation}\label{}
T_{c} ({\bf H})= T_{c}- \frac{1}{2}Q_{ab}H_{a}H_{b}+O (H^{4})
\end{equation}
defines the leading field-dependent anisotropy in the transition
temperature, where $Q_{ab}$ is a tensor describing the coupling of the
the order parameter to the magnetic field. 
The quantity $
\Delta \chi_{ab}=-a (T_{c}-T) Q_{ab} (T)\psi^{2} = \chi_{ab}
(T_{c}^{-})-\chi_{ab} (T_{c}^{+})$ 
is the (reduction) in the magnetic susceptibility tensor associated with the
hidden order.  To explore the non-linear susceptibility in a given
direction $\hat n$ of the magnetic field, we write $H_{a}= H\hat
n_{a}$, so that $T_{c} ({\bf H})= T_{c}-\frac{1}{2}Q (\theta ,\phi )H^{2}$, 
where  $Q (\theta,\phi 
)= \hat  n_{a}Q_{ab}\hat n_{b}$. 
Using thermodynamic arguments, we can explore
the experimental consequences of 
Equation (\ref{fenergy}) \cite{Trinh16}. Solving for $\psi$, we can rewrite
this free energy below $T_c$ as   
\begin{equation}
f[T] = -\frac{a^2}{2b}[T_c ({\bf H})-T]^2.
\end{equation}
so that, taking appropriate derivatives 
in zero field \cite{Trinh16,Ramirez92}, we find that $\Delta
C_{V}/T_{c}=\frac{a^{2}}{b}$, 
\begin{equation}\label{}
\frac{d\Delta \chi_{1}[\theta ,\phi ]}{dT} = \frac{\Delta  C_{V}}{T_{c}}Q[\theta ,\phi ]
\end{equation}
and
\begin{equation}\label{}
\Delta \chi_{3}[\theta ,\phi ]= 3 \frac{a^{2}}{b}Q[\theta ,\phi ]^2
\label{delchi3}
\end{equation}
leading to the 
relationship 
\begin{equation}
\frac{\Delta C_V}{T}\Delta \chi_3 (\theta ,\phi ) = 3\left(\frac{d\chi_1 (\theta ,\phi )}{dT}\right)^2 
\label{thermorel}
\end{equation}
that has been previously checked for \urs along the c-axis \cite{Chandra94}.
Equation (10) holds for {\sl all} orientations of the applied
magnetic field.  From (9) we can estimate  $Q_{zz}$:  
$\Delta \chi_3 = 0.18$ emu/mol T$^{3}$ = $0.18$
 mJ/ mol T$^4$ \cite{Trinh16} (where
we note that 1 emu = 1 mJ/T) and    
$\frac{\Delta C_V}{T_{HO}} = 300$ mJ/mol K$^2$ \cite{Palstra:1985wa}
so we find that 
\begin{equation}
Q_{zz} = \sqrt{\frac{\Delta \chi_3 T_{HO}}{3\Delta C_V}} = .014 \ \mathrm{K/T}^2.
\end{equation}
This value of $Q_{zz}$ suggests that $T_{c}$ vanishes at fields of
$\mu_{0}H_{c}\sim \sqrt{2 T_{HO}/Q}= 50$T, a number
that is roughly consistent with the experimental value \cite{Kim:03}, 
particularly as our extrapolated phase boundary from small fields is expected
to overshoot the measured one. 

In-plane anisotropy has been reported in torque 
magnetometry \cite{Okazaki:2010tn}, cyclotron resonance\cite{Tonegawa:2012}, 
x-ray diffraction \cite{Tonegawa:2014}  and elastoresistivity 
measurements \cite{Riggs:jh} though NMR and NQR studies 
suggest that this nematic signal decreases with 
increasing sample size and also depends on sample quality, suggesting that the bulk is tetragonal \cite{Mito13,Kambe15}.  In principle inter-domain fluctuations
of the basal plane susceptibility contribute to an in-plane $\chi_3$ below $T_c$, and so again experimental consistency with these different measurements 
must be checked. 

For a single domain, broken tetragonal symmetry-breaking manifests
itself through the development of a finite off-diagonal component of
the magnetic susceptibility 
$\chi_{xy}^{D}\sim \left(\frac{V_{D}}{v_{c}} \right) \frac{\langle m_{x}m_{y}\rangle }{T}$,
where $V_{D}$ is the volume of the domain, $v_{c}$ is the volume of a
unit cell and $m_{a}= M_{a}/N_{cells}$ is the magnetization per cell.
The bulk off-diagonal magnetic susceptibility involves an average over
many
different domains that is zero, namely $\overline{\chi_{xy}}=0$
where the over-bar denotes a domain average. 
However domain fluctuations in the susceptibility remain finite,
given by 
\begin{equation}\label{}
\overline{(\Delta \chi_{xy})^{2}}= \left(\frac{V}{V_{D}} \right) (\chi_{xy}^{D})^{2},
\end{equation}
where $V$ is the total volume of the sample. 
The change in the bulk basal-plane nonlinear susceptibility in the hidden
order phase 
is then given by 
\begin{equation}\label{}
\overline{\Delta \chi_{3\perp }}
\sim - \left(\frac{V}{v_{c}}
\right)\frac{\overline{\langle m_{x}m_{y}\rangle^{2}}}{T^{3}}.
\end{equation}
which we rewrite as 
\begin{equation}\label{C2}
\overline{\Delta \chi_{3\perp }}\sim 
- \left(\frac{Vv_{c}}{V_{D}^{2}}
\right)\frac{\left(\chi_{xy}^{D} \right)^{2}}{T}= 
- \left(\frac{v_{c}}{V_{D}}
\right)\frac{\overline{(\Delta \chi_{xy})^{2}}
}{T}
\end{equation}
Thus the inter-domain fluctuations in the symmetry-breaking component 
of the nonlinear susceptibility are expected to generate a
contribution to the basal plane $\Delta \chi_{3\perp}$.

To set bounds on $\Delta \chi_{3\perp}$,
 we compare it with the nonlinear susceptibility along the
z-axis, given by 
$(\chi_{3})_{zzzz}= \left(\frac{V}{v_{c}} \right) \frac{\langle m_{z}^{4}\rangle }{T^{3}}$.
The susceptibility in the z-direction of a single domain is given
by 
$\chi_{zz}^{D} = \frac{V_{D}}{v_{c}} \frac{\langle m_{z}^{2}\rangle}{T}$
so that we can write
\begin{equation}\label{C3}
(\chi_{3})_{zzzz}\sim\left(\frac{V}{v_{c}} \right) \left[\left(\frac{\chi_{zz}^{D}v_{c}}{V_{D}} \right) \right]^{2}\frac{1}{T}
\end{equation}
Taking the ratio of (\ref{C2}) and (\ref{C3}) we obtain 
\begin{equation*}
\frac{
\overline{\Delta \chi_{3\perp }}
}{(\chi_{3})_{zzzz}} 
= -\left(\frac{\chi_{xy}^{D}}{\chi^{D}_{zz}}
\right)^{2}
= -\left(\frac{\chi_{xy}^{D}}{\chi^{D}_{xx}}
\right)^{2}
\left(\frac{\chi_{xx}^{D}}{\chi^{D}_{zz}}
\right)^{2}.
\end{equation*}
We thus see that the magnitude of the anomalous basal plane nonlinear
susceptibility is substantially reduced by the squared ratio of the bulk 
basal-plane and c-axis susceptibilities.

We can rearrange this equation to set bounds on the in-plane
tetragonality 
\begin{equation}\label{}
\left|\frac{\chi_{xy}^{D}}{\chi^{D}_{xx}}
\right| \leq  
\left(\frac{\chi_{zz}}{\chi_{xx}}
\right)
\sqrt{
\left\vert 
\frac{
\overline{\Delta \chi_{3 \perp }}
}{(\chi_{3})_{zzzz}} 
\right\vert .
}
\end{equation}
Putting in numbers, the anisotropy in the linear 
susceptibility is at least five,  
$\left(\frac{\chi_{zz}}{\chi_{xx}}\right) > 5$
while 
the error bounds on the measurement of the in-plane nonlinear
susceptibility are given by 
$\left\vert \frac{\overline{\Delta \chi_{3\perp }}
}{(\chi_{3})_{zzzz}} 
\right\vert  \leq 0.14$
so that 
\begin{equation}\label{}
\left|\frac{\chi_{xy}^{D}}{\chi^{D}_{xx}}
\right| \leq  5 \times \sqrt{0.14}\sim 1.9
\end{equation}
which sets a bound that is two orders of magnitude larger than the
anisotropy measured by torque magnetometry in micron-sized tiny
samples. 
Thus there is no inconsistency between our nonlinear
susceptibility measurement and previous torque magnetometry
measurements. We also see that an order of magnitude improvement in
the nonlinear susceptibility measurements would make it possible to
observe the probe the reported in-plane anisotropy with a bulk measurement.

\section{Open Questions for Future Work}

We next turn to the many open questions, both for experiment and for theory
motivated these angle-dependent susceptibility measurements.

\subsection {Experiment}

Can this angular anisotropy in the hidden order parameter be probed by further
spectroscopic measurements at temperatures in the vicinity of $T_{HO}$?

\begin{itemize}

\item Does the Knight shift display a similar angular anisotropy?

The NMR Knight shift may be a useful additional tool to probe the
g-factor anisotropy at the hidden order transition. 
In \urs the Knight shift closely tracks with the bulk susceptibility
and thus can be used as a cross-check of the angular anisotropy at the
hidden order transition \cite{curro_ref}.  
NMR measurements \cite{Emi:2015ie,Hattori17} suggest that the 
spin contribution to the Knight shift has an anisotropy in
excess of 25. It would be interesting to follow this detailed
anisotropy both as a function of angle and as a function of pressure
in the vicinity of the transition from HO to antiferromagnetism. 


\item Does Raman probe Ising spin fluctuations?

Recent Raman measurements indicate that the most significant
temperature-dependent response is in the $A_{2g}$ channel \cite{Buhot14,Kung15} 
where the measured Raman response function 
\begin{equation}
\chi_{A_{2g}} (\omega,T)
= \int_{o}^{\infty} dt 
\langle O_{A_{2g}}(t), O_{A_{2g}}(0) \rangle 
e^{i\omega t}
\end{equation}
closely resembles the inelastic neutron scattering signal at small 
wavevector \cite{Broholm91}.  Furthermore, the static Raman susceptibility
\begin{equation}
\chi_{A_{2g}}(T) = \frac{2}{\pi} 
\int_0^{25 meV} 
\frac{{\rm Im}\left[ 
\chi_{A_{2g}} (\omega,T)\right]}{\omega} d\omega
\end{equation}
tracks the c-axis magnetic susceptibility \cite{Buhot14,Kung15}.  If we
expand the crystal-field Hamltonian of tetragonal \urs to linear order
in the electromagnetic stress-energy tensor, the
the $A_{2g}$ component of the coupling takes the form 
\begin{equation} 
H = \hat{H}_0 + \hat{O}_{A{2g}} (A_xA'_y - A_yA'_x),
\end{equation}
where unprimed and primed vector potentials refer to in and
outgoing fields, respectively, 
while the operator
\begin{equation}
\hat{O}_{A_{2g}} = \left [ a(\omega) (J_z^2 - J_y^2) J_xJ_y + b(\omega) J_z \right ].
\end{equation}
Here the first term derives from the oscillatory electric field components
($E_{x}E_{y}'-E_{y}E_{x}'$) of the stress-energy tensor 
while the second term derives from the Poynting vector $\hat  z\cdot (\vec{E}\times  \vec{B}' +\vec{ E}'\times \vec{ B})$. The close resemblance between the Raman signal and the measured spin 
fluctuations \cite{Buhot14,Kung15} 
suggests that this second magnetic $J_z$ term is dominant. More work is needed
to determine the relative importance of $a(\omega)$ and $b(\omega)$, particularly for strongly spin-orbit coupled materials.

\end{itemize}

\subsection{Theory}

Some argue that the hidden order parameter is elusive because it is fundamentally complex.  In this approach a performed band of Ising quasiparticles with half-integer angular momentum form a multipolar density wave.  However because \urs
is tetragonal, $J_z$ is conserved (mod 4).  This angular momentum exchange of
$\pm 4 \hbar$ implies mixing of states, for example of the form 
\begin{equation}
|\bold{k} \pm \rangle = \alpha |\bold{k},\pm\frac{5}{2}\rangle + 
\beta |\bold{k},\mp \frac{3}{2} \rangle
\end{equation}
that will lead to a finite transverse 
coupling ($\Phi \propto |\alpha \beta|^2$) that is ruled out by the observed 
Ising anisotropy observed in experiment.  How to reconcile this approach with
experiment?

Another tack is to argue that the hidden order parameter is elusive because
it is a fundamentally novel nonlocal order parameter as occurs in 
superconductivity \cite{Hansson04}.
In particular it could be the case of a fractionalized
order parameter, for example the square root of a 
multipole.  
One such proposal \cite{Chandra:2013gv,Chandra14,Chandra15} 
suggests that the itinerant f quasiparticles have integer angular momentum
due to a coherent, symmetry-breaking 
hybridization of the conduction electrons with 
integer spin f-states.  In this case the Ising anisotropy is preserved since
the up- and down- spin configurations differ by at least two units of 
angular momentum.  This approach predicted \cite{Chandra:2013gv,Chandra14,
Chandra15} 
\begin{equation}
\Delta \chi_3 \propto \cos^4 \theta
\end{equation}
but the microscopic theory needs revision, partially due to the absence of 
the predicted transverse moment. There we treated the hybridization of
$f$-moments with a simplified s-conduction band; we now know, due to interest
in topological Kondo insulators, that with this approach $SmB_6$ is a metal.  
Instead it is crucial that we consider p-wave hybridization \cite{Duan97} and 
this will surely 
affect the microscopics and the gap structure leading to new, verifiable
predictions for experiment.

We acknowledge stimulating exchanges with N. Curro, N. Harrison and T. Hattori. 
This work was supported by National Science Foundation grants
NSF DGE 1339067 (J. Trinh), NSF DMR-1334428 (P. Chandra), NSF DMR 1309929 
(P. Coleman), NSF DMR 1534741 (A.P. Ramirez) and Ames Laboratory
Royalty Funds and Iowa State University startup funds (R. Flint). The Ames Laboratory is operated for the U.S. Department of Energy by Iowa State University under Contract No. DE-AC02-07CH11358.

%

\end{document}